\documentclass[conference]{IEEEtran}
\IEEEoverridecommandlockouts

\usepackage{adjustbox}
\usepackage{textcomp}
\usepackage{graphicx}
\usepackage{xcolor}
\usepackage{soul}
\usepackage{comment}

\usepackage{array, caption, floatrow, tabularx, makecell, booktabs}
\definecolor{mygreen}{RGB}{102,240,80}

\usepackage{cite}
\usepackage[acronym,shortcuts]{glossaries}
\usepackage{amsmath,amssymb,amsfonts}
\usepackage{algorithmic}
\usepackage{graphicx}
\usepackage{tikz}
\usepackage{textcomp}
\usepackage{xcolor}
\usepackage{siunitx}
\def\BibTeX{{\rm B\kern-.05em{\sc i\kern-.025em b}\kern-.08em
    T\kern-.1667em\lower.7ex\hbox{E}\kern-.125emX}}
\usepackage{array,multirow}
\usepackage{hyperref}
\newcommand{\parag}[1]{\noindent\textbf{#1. }}
\begin{document}

\title{Penetration Testing of 5G Core Network \\ Web Technologies}

\author{\IEEEauthorblockN{Filippo Giambartolomei}
\IEEEauthorblockA{\textit{Ikerlan Technology Research Centre}, \\
Basque Research and Technology Alliance (BRTA) \\
Arrasate-Mondragón, Spain\\
fgiambartolomei@ikerlan.es}
\and
\IEEEauthorblockN{Marc Barceló}
\IEEEauthorblockA{\textit{Ikerlan Technology Research Centre}, \\
Basque Research and Technology Alliance (BRTA) \\
Arrasate-Mondragón, Spain\\
mbarcelo@ikerlan.es}
\and
\IEEEauthorblockN{Alessandro Brighente}
\IEEEauthorblockA{\textit{Deparments of Mathematics} \\
\textit{University of Padua}\\
Padua, Italy\\ 
alessandro.brighente@unipd.it}
\and
\IEEEauthorblockN{Aitor Urbieta}
\IEEEauthorblockA{\textit{Ikerlan Technology Research Centre},\\
Basque Research and Technology Alliance (BRTA) \\
Arrasate-Mondragón, Spain\\
 aurbieta@ikerlan.es}
\and
\IEEEauthorblockN{Mauro Conti}
\IEEEauthorblockA{\textit{Deparments of Mathematics} \\
\textit{University of Padua}\\
Padua, Italy\\
mauro.conti@unipd.it
}
}

\maketitle
\begin{tikzpicture}[remember picture, overlay]
  \node[font=\sffamily\normalsize, yshift=-0.7cm, text centered, text width=\paperwidth, anchor=north west] at (current page.north west) {%
This paper has been accepted for publication at IEEE International Conference on Communications (ICC) 2024 %\url{TODO}
  };
\end{tikzpicture}
\begin{abstract}
% The Fifth Generation (5G) of mobile networks provides increased capabilities compared to previous generations thanks to its high virtualization. Indeed, 
Thanks to technologies such as virtual network function the Fifth Generation (5G) of mobile networks dynamically allocate resources to different types of users in an on-demand fashion. Virtualization extends up to the 5G core, where software-defined networks and network slicing implement a customizable environment. These technologies can be controlled via application programming interfaces and web technologies, inheriting hence their security risks and settings. An attacker exploiting vulnerable implementations of the 5G core may gain privileged control of the network assets and disrupt its availability. However, there is currently no security assessment of the web security of the 5G core network.

In this paper, we present the first security assessment of the 5G core from a web security perspective. We use the STRIDE threat modeling approach to define a complete list of possible threat vectors and associated attacks. Thanks to a suite of security testing tools, we cover all of these threats and test the security of the 5G core. In particular, we test the three most relevant open-source 5G core implementations, i.e., Open5GS, Free5Gc, and OpenAirInterface. Our analysis shows that all these cores are vulnerable to at least two of our identified attack vectors, demanding increased security measures in the development of future 5G core networks.
% The rapid expansion of 5G networks into various domains, including mobile, automotive, Internet of Things (IoT), and potentially scheduled flights, requires a focus on the security of the 5G Core network. This paper specifically analyzes the security of the 5G Core through a penetration testing approach, aiming to identify weaknesses in deployed 5G Core implementations. The study compares three open source 5G Core solutions: Open5gs, Free5gs, and OpenAirInterface, evaluating their vulnerabilities and proposing countermeasures. The findings serve as a starting point for future research and contribute to the emerging field of cybersecurity in 5G Core networks, encouraging further exploration in this critical domain.
\end{abstract}


\section{Introduction}
% 1) parte su perchè ci serve 5G e perchè è meglio delle generazioni precedenti (trovare un po' di numeri data rate raggiunti, copeartura,..) e cita survey su 5G ed evoluzione 

% 2) focus su 5G core, cos'è e a cosa serve. Focus sulla sicurezza del core e impatto di attacchi ad esso. 

% 3) spiegare chi ha fatto lavori del genere in letteratura e cosa manca, ovvero cosa facciamo noi in più. 

% 4) spiegare che facciamo anslisi sicurezza 5G core in generale facendo riferimento a standard. Dire che utilizziamo delle soluzioni open source e spiegare perchè sono valide e di conseguenza ci danno info utili su 5G core in generale 

% 5) our contributions can be summarized as follows: elenco puntato 
The 5G core is the heart of a 5G network. Its main roles include providing end users with secure and reliable connection and access to a set of services. Although the network core's role did not change with respect to previous generations, its implementation has been completely revolutionized. Indeed, all 5G core network functions are completely software-based and designed as cloud-native \cite{etsiCore}. These changes were necessary to accommodate the increasing number of connected users and their different requirements (based on traffic type). Indeed, the performance that 5G networks are expected to meet, includes a remarkable data rate requirement of 20Gbps for download and 10Gbps for upload. This represents a substantial leap forward when compared to the capabilities of 4G/LTE Release 14, which provides a maximum data rate of 1Gbps for downlink and 50Mbps for uplink \cite{WEBSITE:10}. 2019 marked the beginning of the deployment and spreading of the fifth generation network, more commonly known as the 5G network  \cite{WEBSITE:8}. 

The development of a new generation is not solely driven by the increased performance demand for communications, but also by an increased need for security and privacy guarantees. Indeed, the 5G core is responsible for establishing secure connections and providing basic services such as authentication and authorization \cite{tang2022systematic}. However, although security improved compared to the previous generations, there still exist security and privacy gaps that may jeopardize 5G networks \cite{ID:5}. Recent examples showed that it is possible to create fake base station attacks \cite{lotto2023baron}, or to tamper with the management of the identifiers \cite{chlosta20215g}. Therefore, a thorough assessment of the security of 5G core network implementation is a fundamental need to develop secure networks. 

Despite the importance of the issue, the literature on this topic is still scarce, providing only a few examples of the analysis of the 5G core security. Holtrup et al. \cite{ID:6} provide a formal analysis of the security of the 5G network, considering both stand-alone and non-stand-alone implementations. However, the authors do not follow a standardized methodology for security assessment. Pell et al. present a more specific approach bearing in mind the MITRE Threat modeling framework \cite{ID:7}. However, such an approach is not sufficient to clearly understand the broader class of attack types and it is particularly dependent on the system implementation and/or version. Altariqi et al. \cite{ID:1} consider the security of the 5G core, however focusing only on its function virtualization capabilities. 
%While, Marian Gusatu and Ruxandra F. Olimid shows some mitigation for DDoS attacks regarding a specific component of the 5G Network: the Multi-access Edge Computing \cite{ID:8}. 
% 5G Core, indeed, implements a totally virtualized environment, enhancing the portability and usability of networking services. Considering the rapid expansion in the number of appliances, even the bandwidth had to keep up. Specifically for this reason, the download and upload speed boosted significantly, and side by side with these improvements, also the virtualization of the functionalities migrated from physical world to a virtual one, with respect to 4G. This change gives the possibility to every individual to set up its own 5G network with all the benefits it brings.

% The main reasoning for the overwhelming of his predecessor 4G, diffused starting from the 2009, was the massive growth of the number of devices connected to the internet and, as a result, the huge increase of traffic.
% The main reason behind the spreading of 5G network is the virtualization of every function, leading to better flexibility and toughness in case of errors. 

% In the last years, a lot of papers related to this topic have been performed and published, mainly about how 5G network works. Whilst, for what concerns the security of the 5G, and especially regarding a crucial component such as the 5G Core, the material and consideration expressed are really scarce. 

In this paper, we present the first security assessment of the 5G core from a web security perspective. We leverage STRIDE \cite{shostack2008experiences}, a well-established threat modeling methodology developed at Microsoft which is one of the de facto standard approaches. While drawing inspiration from Altariqi et al. work \cite{ID:1}, our paper takes a different approach by focusing on the web components of the 5G core. To assess the security of the 5G core network, we test our model on three public and well-established 5G core implementations, namely Open5GS\footnote{https://open5gs.org/}, Free5GC\footnote{https://free5gc.org/}, and OpenAirInterface\footnote{https://openairinterface.org/}. We leverage a suite of penetration testing tools to implement a large set of attacks and identify a relevant number of attacks that can be performed on these 5G core implementations.

The contributions of our paper can be summarized as follows:
\begin{itemize}
    \item We propose the first web-based threat model for the 5G core. We leverage the STRIDE methodology, a well-known approach to systematically derive all possible threat vectors affecting the 5G core. For each threat, we discuss how a successful attack could impact the 5G network.
    \item We test the web security of multiple well-known implementations of the 5G core. Being these implementations compliant with the 5G standard and used by many companies and research institutions, they represent what is closest to a real 5G core implementation. Via our testing, we point out that all of these cores are vulnerable to at least two attacks, thus requiring cautious deployment and further security measures.
\end{itemize}

The rest of the paper is organized as follows. We introduce the system and threat models in Section II. In Section III, we present the set of 5G core implementations that we tested and the set of security tools used to cover all the threats we identified in Section II. We then present the results. In Section IV we present possible countermeasures to the identified attacks and then derive the conclusions in Section V.
%\section{Related Works}
%\section{Background}
\begin{comment}
    \begin{figure}[!ht]
    \centering
    \captionsetup{justification=centering}
    %\includegraphics[scale=0.18]{images/Open5GS.jpg}
    \includegraphics[width=\textwidth]{images/Open5GS.jpg}
    \caption{Overview of the 5G Core architecture (Open5gs).\hl{figure should appear after it has been introduced in the text. Also, good to make it bigger for readibility}}
    \label{fig:open5gs}
\end{figure}
\end{comment}

%in order to
%Similar to 
%For instance
%These scenarios
%As a result
%Apart from that
%The remainder of the paper
%Focusing the attention
%The reason behind
%It is commonly used
%The basic concept is

\section{System and Threat Model}
We provide a description of the considered system and threat model. In Section~\ref{sec:sys}, we present the components of the 5G core network focusing on their use of web-based technologies. We then present in Section~\ref{sec:thMod} the STRIDE-based threat model for the 5G core.

\subsection{System Model}\label{sec:sys}
% descrizione high level del 5G core da standard

A native 5G core implements a service-based architecture thanks to which it is possible to independently scale different functions~\cite{brown2017service}. Evolved Packet Core (EPC) functions are instead based on a point-to-point architecture and have overlapping responsibilities. The integration of new functions happens via a message bus for control plane interactions. It implements (REST)-ful application programming interfaces (APIs) that use hyper-text transport protocol (HTTP/2) inquiries for high-level flexibility~\cite{mayer2018restful}. Compared to previous generations, 5G implements a totally virtualized environment, enhancing the portability and usability of networking systems and services. In particular, the 5G core extensively uses virtualization techniques to implement Virtual Network Functions (VNF), network slices, and software-defined networking to implement a scalable and customizable network of services.

\begin{figure*}[!t]
    \centering
    \includegraphics[width=.8\columnwidth]{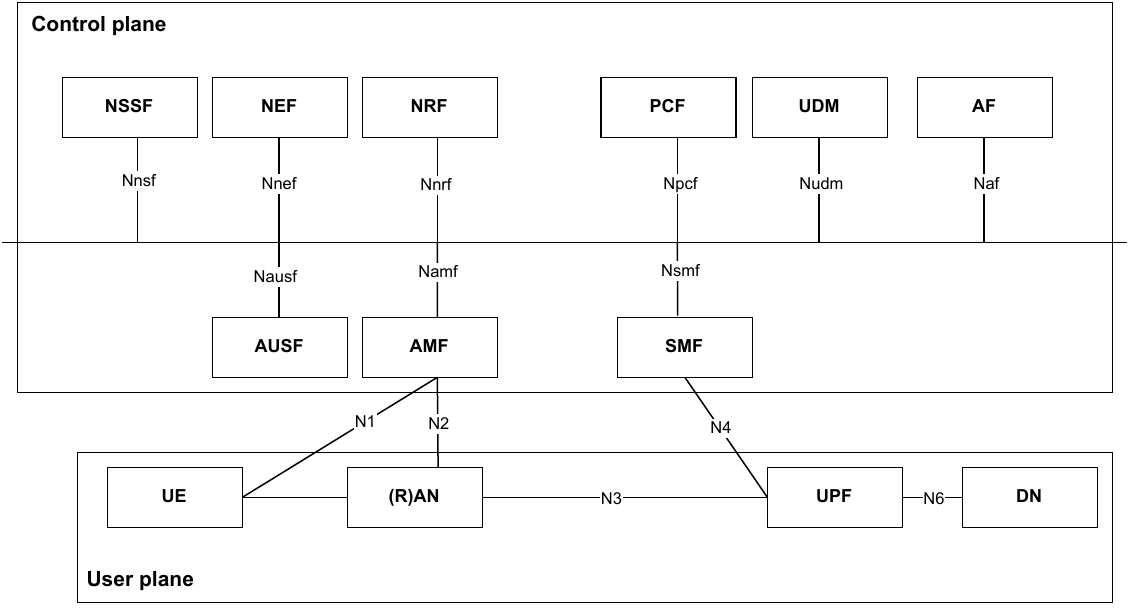}
    \caption{Overview of the 5G Core architecture.}
    \label{fig:5gcore}
\end{figure*}

Fig.~\ref{fig:5gcore} shows the standard implementation of the 5G core \cite{etsiCore}. The major components are:
\begin{itemize}
    \item Access and Mobility Management Function (AMF): responsible for access control and mobility, and includes Network Slice Selection Function (NSSF).
    \item Session Management Function (SMF): manages sessions according to network policies.
    \item User Plane Function (UPF): responsible for packet-level operations, such as packet routing, forwarding, and inspection.
    \item Policy Control Function (PCF): provides a policy framework for network slicing, roaming, and mobility management.
    \item Unified Data Management (UDM): stores subscriber data and profiles.
    \item Network Function Responsibility Function (NRF): provide registration and discovery functionalities such that network functions can communicate with one another via APIs. 
    \item Network Exposure Function (NEF): API gateway that allows external users the ability to monitor, provision, and enforce application policies for users inside the operator network.
    \item Authentication Server Function (AUSF): authentication server.
    
\end{itemize}

All these components are connected via standard interfaces (e.g., N1 and N2) defined by 3GPP. To manage and control network functions at different levels, both operators and services use API calls with a request-response mechanism passing via \textit{reference points}. 
%Until now, the NFs have been considered as isolated functionalities that matters for the correct operation of the 5G Core and in general 5G network. What has not been mentioned, is the fact that the Network Functions communicate each other using reference points, thus, a point to point connection. These reference points are, indeed, nodes from which NFs connect to other NFs. 
Each of these reference points can be enforced with the use of Service Based Interfaces (SBI) whose usage is based on a logically shared framework between functions. One meaningful consideration that must be done is that SBIs are mainly used inside the Control Plane, indeed, the User Data Plane does not present SBI, but it only owns the reference points. 

% This is possible due to the use of the network slicing technique \cite{WEBSITE:9}, a fundamental mechanism used for the management of multiple virtual networks over a single physical network infrastructure. From a high-level point of view, we can divide the 5G Core into two main technologies: the Software Defined Network (SDN) \cite{ID:3} which brings a lot of benefits to 5G networks splitting the User Data plane from the Control plane; and then the Network Functions Virtualization (NFV) \cite{ID:4}. The latter, merged together with the SDN comprises all the functionalities of the network and how the 5G Core works. At the beginning, the 5G network was deployed in Non-Standalone (NSA) mode, with the help of the previous generation's infrastructure. Even if the Standalone (SA) option has already started to be implemented, this process is still in progress \cite{WEBSITE:7}. Apart from the deployment of this hybrid structure, digging a bit deeper inside the 5G Core, we can notice how its architecture is based on micro services that implement the so-called Service Based Architecture (SBA). The latter provides a comprehensive framework that handles all the main functionalities used in 5G networks. The principal operation of the SBA is handling Network Functions (NFs): AMF, SMF, PCF, AUSF, UDM, NSSF, NEF, NRF, AF, UPF and others; each of these providing services to the other functions and users, and each one with a specific task to accomplish. 

This huge increment of virtualization within the 5G core infrastructure engenders an urgent requirement for a graphical user interface (GUI). This imperative GUI serves as an essential tool that facilitates administrators in the seamless and rapid management of subscriber connections. Through this graphical interface, administrators have the chance to oversee, control, and optimize the connectivity of subscribers, thereby enhancing the overall efficiency and efficacy of the 5G core network. Furthermore, the network requires storage capabilities to account for the different configurations and data at rest about users and operators. All these functionalities can be enabled via web-based technologies.

\subsection{Threat Model}\label{sec:thMod}
% descrizione dell'attaccante e delle sue capacità. STRIDE thing

% The section aims to explain what a threat model is and what threat categorization has been chosen in order to exploit the attacks taken into account (Tab. \ref{table:STRIDE}). A threat model can be considered as an organized procedure which helps to assess the security hazards of an application, a system or whatever: in our case the 5G Core network. More specifically, the idea is to show which resources the system is trying to protect, with a deep inclination on how to avoid threats and how to mitigate them. For our purpose we choose to employ the STRIDE categorization, thoroughly explained in the next section.\\

%The 5G Core, as already mentioned, is composed by different constituents, each of these broadens the attack surface and the possibility of flaws. In this paper we consider a powerful attacker having access to the 5G core. We put no restriction on the attacker's capabilities, nor do we consider it to have limited resources. Indeed, attackers may aim to exploit vulnerabilities in web interfaces to gain unauthorized access, manipulate network settings, or disrupt network operations. 

We consider a powerful attacker that has a large amount of computation power and resources to try accessing the 5G core. We put no restriction on the attacker's capabilities to expose a large set of possible threats. 
We leverage the STRIDE threat modeling approach \cite{shostack2008experiences} to identify vulnerabilities in the 5G core. STRIDE is an acronym that encompasses the main threats that might affect the system as visible in Tab.~\ref{table:ThreatModel}, i.e.,  
\begin{itemize}
    \item \textbf{Spoofing:} a malicious user impersonates a legitimate user, deceiving the receiver that the communication is lawful.
    \item \textbf{Tampering:} a baleful user alters or modifies legitimate data such that they are harmful to the receiver. Commonly paired with the spoofing technique.
    \item \textbf{Repudiation:} A malicious actor cannot be unambiguously linked to an action/transmission that happened.
    \item \textbf{Information disclosure:} it is the ability to gather sensitive system information due to unprotected data flows.
    \item \textbf{Denial of service:} a baleful user disrupts resource availability by overwhelming a communication link with packets/requests.
    \item \textbf{Elevation of privilege:} the attacker gains access to information that a regular user should not be authorized to access.
\end{itemize}
Tab.~\ref{table:ThreatModel} provides a list of possible attacks, a list of possible attack implementations, and the attacker's goals for each of the threats identified by STRIDE. We assume the attacker uses these attacks to find and exploit vulnerabilities in the 5G core.
In the following, we test the different 5G core implementations encompassing all the threats listed by the STRIDE methodology. We note that, although not providing a full list of all possible attacks, our identified attacks cover the OWASP top 10 web application security risks~\cite{owasp}.

\begin{table*}[!t]
\addtolength{\topmargin}{+0.15cm}
\caption{Potential threats for STRIDE modeling.}
    \label{table:ThreatModel}
%\resizebox{\textwidth}{!}{%
\begin{tabular}{|p{3.5cm}|p{3.5cm}|p{4.5cm}|p{4.5cm}|}
\hline
\textbf{Threat} &
  \textbf{Possible attacks} &
  \textbf{Possible implementations} &
  \textbf{Attacker's goal} \\ \hline  \hline
Spoofing &
  Bruteforce attack, Dictionary attack. &
  Exploits the absence of/weak encryption,  to hijack the traffic and modify it according to necessity, leading to identity theft or MiTM attack. &
  Intercept the traffic with the purpose of impersonation, sharing false information or obtaining sensitive information. \\ \hline
Tampering &
  Bruteforce attack, Dictionary attack, SQL injection, NoSQL injection. &
  Usually coupled with the spoofing technique, allows to modify the data traffic to deceive the user or obtain useful information like user credentials. &
  Make the victim unaware that the data traffic has been modified. \\ \hline
Repudiation &
  JWT Robustness, Clickjacking. &
  Possibility to modify content pretending to be another user, i.e., Identity Theft. &
 Falsify data, and alter transmission, with the intent of deceiving or defrauding others without being caught. \\ \hline
Information disclosure &
  Permission leakage, SQL injection, NoSQL injection, Directory traversal. &
  The malicious actor may gain unauthorized access to a list of usernames, passwords, and even data configurations. &
  Acquire, alter, and utilize data to which the attacker should not have legitimate access. \\ \hline
Denial of service &
  DoS and DDoS attack. &
  As an illustrative instance, a malicious actor could render the Access and Mobility Management  Function (AMF) unavailable, thereby making unsuitable the users' administration. &
  Capability to render a service inoperable, causing it to crash, ceasing functioning, or potentially slowing down its operational efficiency.\\ \hline
Elevation of privileges &
  JWT Robustness,   Bruteforce attack, Dictionary attack.  &
  The potential to assume the identity of an administrator, to gain more privileges and thereby undertaking actions beyond those initially authorized. &
  Seeking to elevate his privileges, thereby enabling him to perform actions that were, until that moment, beyond his authorized capabilities. \\ \hline
\end{tabular}%
%}
\end{table*}

\section{Tests implementation and results}
% 1) chosen 5G core
% 2) tools and why choosing them
% 3) analisi degli attacchi e risultati

\subsection{Chosen 5G Core}
Our objective of studying the web security of the 5G core requires a proper implementation of the overall standard defined by the 3GPP \cite{etsiCore}. However, implementing it from scratch does not provide significant insights into currently available deployments and solutions.
To assess whether state-of-the-art implementations of the 5G are vulnerable to the threats identified via the STRIDE methodology from a web-based standpoint, we refer to the Open5GS, Free5gc, and OpenAirInterface implementations. In other studies related to threats associated with the 5G core, Neto et al. used a similar setting, considering Magma rather than OpenAirInterface \cite{ID:2}. Moreover, at the current time, these three parsed cores are the most cutting-edge in the marketplace, used both in research and industry for research and development~\cite{liu2022evaluation,kim2022implementation,mamushiane2023deploying,dumitru2022analysis,nikaein2014openairinterface}.

\subsection{Tools}
To test all the threats in STRIDE, we leverage a set of well-known and established security tools. We choose them considering the performance they provide and the results that can be obtained using them. In the following, we provide a brief description of the tools and their main scope:

\parag{Nmap}
This tool is mainly employed to gather information about a specific network or host. Nmap, indeed, allows the user to scan one or more targets simultaneously, to obtain as much information as possible regarding services, ports, hosts, and vulnerabilities.

\parag{WhatWeb}
Its purpose is primarily to
identify all the technologies and features the system presents from a web perspective. It can identify entities, packages, and servers in a way similar to what nmap does for networks.

\parag{Nikto}
This web application scanner is mainly employed to get basic information about servers, checking for outdated system versions, unusual headers, and everything that can be exploited against the target.

\parag{Dirbuster and Dirb}
Both the mentioned web content scanners are employed to brute force a specific web page, to discover hidden directories and hidden files. The attacker can use these tools to identify all the directories available to the web architecture.

\parag{Burpsuite}
This penetration testing tool is one of the most employed by the community. It provides a large variety of techniques implementable for several attacks.

\parag{Nessus}
This tool is a powerful vulnerability scanner. This allows to inspect the objective from a general perspective but also delving into its details providing specific scan templates, focusing only on specific ports, services or login pages.

\parag{Hping3}
It is mainly employed for the creation of ICMP, TCP and UDP packets. This network tool has the purpose of crafting, following some specifications, the most effective packet; such as a DoS attack. 

\parag{SlowLoris}
Also known as Low bandwidth DoS tool allows to exploit a Denial of Service (DoS) against a targeted web server. This tool employs the "low and slow" technique, opening connections with the server and keeping them open.

\parag{Hulk}
HTTP Unbearable Load King DDoS is a tool used to assemble and deliver unique pattern packets.

\parag{JohnTheRipper}
This tool has the leading purpose of cracking passwords. 

\subsection{Results}

\parag{Permissions leakage, SQL and NoSQL injections}
Open5gs and OpenAirInterface are both affected by this exposure, even if in a distinct fashion. The first, as a result, allows an attacker to remotely access the MongoDB core's database without any kind of authentication, granting to look, add, modify, and even remove accounts. While, for the second core, the approach was a little disparate, especially since OpenAirInterface does not own an unstructured database. Therefore, at implementation time, the latter holds two users: the "root" which possesses all the privileges, and the "user" which owns the same permissions as the administrator, and it should not. Even if accessing a different kind of DB, the weakness exploitation is pretty much the same as the Open5gs. In both cases, the outflow of privileges leads to injections: whether SQL or NoSQL. In addition, in both cases, we discovered the Principle of Least Privilege \cite{WEBSITE:5} (PoLP) was not respected, and neither was the absence of user input sanitization. In the matter of Free5gc, this core was not vulnerable to information leakage and therefore to injection of any type, mainly because it does not enable external users to remotely access its database.

\parag{Dictionary and Bruteforce attacks}
These attacks cannot be tested in the OpenAirInterface because of the absence of the web interface, and therefore these results as not implementable. In such a way, as visible from the table in Tab.~\ref{table:compTable}, this offense has succeeded for the Open5gs, whilst it did not succeed for the Free5gc. The latter indeed, presents hard-coded credentials, preserving the login details inside a stored file. Even if this is not a good practice, it did not allow to brute force the credentials. Moreover, it possesses NIA2: an integrity protection algorithm that does not allow users to read or modify the content of the packets without a specific key.

\parag{DoS and DDoS}
In the matter of Open5gs, this Core showed a little delay during the opening of the page, but nothing remarkable intended to be considered as vulnerability. Similarly, OpenAirInterface did not seem to be affected by this kind of attack, it indeed kept regularly working, even if for this specific instance the attack has been implemented against some of its containers. Nevertheless, Free5gc appeared exposed unlike the other two. When testing the latter, it revealed a significant delay during the opening of the login page. This isolated scenario turns out to be vulnerable to DDoS attack, exploited using the \texttt{Hping3} tool in flooding mode, randomizing the source IPs of each request. Although the web server did not crush, this offense may be deemed as effective also considering a large-scale uptake.  

\parag{Directory traversal}
For this particular attack, only Free5gc was vulnerable, indeed, the examined folder discovered without knowing its specific path was \texttt{/etc/passwd} enabling in this circumstance the leakage of a large amount of critical information that can be exploited by malicious actors.

\parag{Clickjacking}
About OpenAirInterface, here the attack is not feasible, due to the absence of a web interface. Clickjacking seemed to be a feasible attack, taking into account that either Open5gs and Free5gc enable \texttt{IFRAME}. This was not sufficient, indeed, we did not reach the expected result, mainly due to the presence of the Cross-Origin Resource Sharing policy (CORS) which forestalls to access web resources through different sources than the ones employed by the website. 

\parag{JSON Web Token}
Open5gs turned out to be vulnerable to this attack because it uses a known passphrase. Therefore, once obtained the key, it is possible to change the content of the JWT's payload and then recreate the proper signature through the previously obtained key, making it valid. Regarding Free5gc, it does not rely on cookies, or hard-coding the credentials, thus it does not save the logged user's session. Whilst, for the OpenAirInterface, this category of offense is not even applicable, because of the absence of a web interface.

Table \ref{table:compTable} presents a summary of the attacks tested against the different 5G core implementations, together with the effectiveness outcome. We notice that not all cores are vulnerable to the same attacks, and for some of them, attacks were not implementable at all.

\begin{table}[h!]
\caption{Comparison Table between tested attacks over the three mentioned 5G Core.}
    \label{table:compTable}
\resizebox{\columnwidth}{!}{
\begin{tabular}{|l|l|l|ll|}
\hline
\textbf{Attacks}                                            & \textbf{Open5Gs} & \textbf{Free5Gc} & \multicolumn{2}{l|}{\textbf{OpenAirInterface}} \\ \hline \hline
{\color[HTML]{000000} 
\textbf{Implementable attacks}} & 9              & 9               & \multicolumn{2}{l|}{4} 
\\ \hline
{\color[HTML]{000000} 
\textbf{Successful attacks}} & 5              & 2               & \multicolumn{2}{l|}{2} 
\\ \hline \hline
{\color[HTML]{000000} 
\textbf{Database permission leakage}} & yes              & no               & \multicolumn{2}{l|}{yes}                       \\ \hline
{\color[HTML]{000000} \textbf{SQL injection}}               & no               & no               & \multicolumn{2}{l|}{yes}                       \\ \hline
{\color[HTML]{000000} \textbf{NoSQL injection}}             & yes              & no               & \multicolumn{2}{l|}{no}                        \\ \hline
{\color[HTML]{000000} \textbf{Dictionary attack}}           & yes              & no               & \multicolumn{2}{l|}{not implementable}                         \\ \hline
{\color[HTML]{000000} \textbf{Bruteforce attack}}           & yes              & no               & \multicolumn{2}{l|}{not implementable}                         \\ \hline
{\color[HTML]{000000} \textbf{DoS and DDoS}}                & no               & yes              & \multicolumn{2}{l|}{no}                        \\ \hline
\textbf{Directory traversal}                                & no               & yes              & \multicolumn{2}{l|}{not implementable}                         \\ \hline
\textbf{Clickjacking}                                       & no               & no               & \multicolumn{2}{l|}{not implementable}                         \\ \hline
\textbf{Json Web Token robustness}                          & yes              & no               & \multicolumn{2}{l|}{not implementable}                         \\ \hline
\end{tabular}
}
\end{table}

% Figure \ref{fig:graph} summarizes the number of attacks attempted against each 5G core implementation together with the number of successful attacks.
% \begin{figure}[H]
%     \centering
%     \captionsetup{justification=centering}
%     \includegraphics[scale=0.28]{images/GraficoABarre.pdf}
%     \caption{Outcomes among tested attacks over 5G Cores.}
%     \label{fig:graph}
% \end{figure}

\section{Countermeasures}
This section involves all the possible expedients and countermeasures to avoid the attacks formerly mentioned and exploited. The presented precautions are some of the most commonly employed, nevertheless not unique. Starting from the former section introduced earlier, the following list presents some possible tricks to avert \textit{Permission leakage, SQL and NoSQL injection:} 
\begin{itemize}
    \item The first method is to apply the principle of least privilege (PoLP) in which each user possesses the minimum levels of privileges.
    \item The second regards the compulsoriness of the authentication for remote connections.
    \item The third approach, instead, is inherent to the input sanitization: which means removing unsolicited characters from the input, before processing them.
\end{itemize}
The concept that stands behind each of these tricks is to not supply more permissions to users than the ones needed.

%\begin{comment}

%\end{comment}

% \begin{figure}[!ht]
%     \centering
%     \captionsetup{justification=centering}
%     \includegraphics[scale=0.49]{images/Table.png}
%     \caption{Comparison Table between tested attacks over three 5G Cores.}
%     \label{fig:compTable}
% \end{figure}

% \textbf{Legend:}
% \begin{itemize}
%     \item \textcolor{mygreen}{Yes} = Exploitable
%     \item \textcolor{red}{No} = Not Exploitable
%     \item \textcolor{blue}{/} = Not applicable
% \end{itemize}

Taking a step forward, pursuing with the \textit{Dictionary and Bruteforce attacks}, the best way to prevent both, as visible from the failure achieved with Free5gc, is to attach some encryption algorithm.
In such a way, the packets can be captured but not read or modified. 

In addition to these scenarios, the tested \textit{DoS and DDoS} exhibit several countermeasures even if there is no assurance of being 100\% safe:
\begin{itemize}
    \item The firewall is the first methodology that can be applied with its editable rules.
    \item Intrusion Detection System (IDS) is another significant system to avert DoS or DDoS, primarily employed to monitor the network, trying to understand if an attack is ongoing.
    \item Introducing a connection timeout or even limiting incoming data rate (SlowLoris).
    \item Extending the backlog of pending connections (SlowLoris).
    
\end{itemize}

Now, moving on to the fourth attack, known as \textit{Directory Traversal}. This vulnerability hinges on two crucial concepts that should be considered to prevent exploitation. It's important to note that Free5gc, unfortunately, lacks of input sanitization, as evidenced by the exploit ../../. Therefore, it is essential to thoroughly analyze and carefully process any input characters to safeguard against this vulnerability. \cite{WEBSITE:6}. Moreover, another significant expedient that we can employ to avoid the Directory traversal is the already mentioned PoLP.

Concerning the \textit{Clickjacking} attack, we will not delve too much into details mainly because this attempt wasn't successful and because it was already presented with an effective countermeasure such as the CORS policy.

Taking a step forward to the last selected attack, there are two main possibilities to avoid its misuse. To avert the exploitation of the \textit{Json Web Token}, one possible way is to avail of an uncommon password; the latter indeed, should not be present in any of the known wordlists. Moreover, another potential countermeasure should be to decrease the IAT (issued at) that identifies how long the JWT is alive. The concept is to employ an adequate amount of "time to live" based on the usage of the token.

\section{Conclusions}
Form Table \ref{table:compTable}) we observe that among the proven attacks, there were no offenses that affected every Core in the same manner. This mainly arises from the fact that, even if considering the same standard core, every single implementation exhibits a small degree of changes that characterize the structure it deploys. The purpose of this paper is to give insights regarding which one of the analyzed three open sources Core is the best in terms of security from a web technology perspective. As already mentioned there is a consideration that must be done, mainly due to the inability to test these components in the same manner. Indeed, for OpenAirInterface we were not able to prove some of the already mentioned exploits attempted with the other cores. This leads to the conclusion that the Free5Gc is the most secure of the three analyzed Cores and over the tested attacks. However, this is an important point because each of the leveraged exploits presents a severity that must be taken into account; since not all the flaws found have the same criticality. 

\section*{Acknowledgments}

The European commission financially supported this work through Horizon Europe program under the IDUNN project (grant agreement number 101021911).

\bibliographystyle{IEEEtran}
\bibliography{bibliography} 

\end{document}